# QuIP: A P4 Quantum Internet Protocol Prototyping Framework

Wojciech Kozlowski, Fernando A. Kuipers, *Senior Member, IEEE*, Rob Smets, and Belma Turkovic

*Abstract*— Quantum entanglement is so fundamentally different from a network packet that several quantum network stacks have been proposed; one of which has even been experimentally demonstrated. Several simulators have also been developed to make up for limited hardware availability, and which facilitate the design and evaluation of quantum network protocols. However, the lack of shared tooling and community-agreed node architectures has resulted in protocol implementations that are tightly coupled to their simulators. Besides limiting their reusability between different simulators, it also makes building upon prior results and simulations difficult. To address this problem, we have developed QuIP: a P4-based Quantum Internet Protocol prototyping framework for quantum network protocol design. QuIP is a framework for designing and implementing quantum network protocols in a platform-agnostic fashion. It achieves this by providing the means to flexibly, but rigorously, define device architectures against which quantum network protocols can be implemented in the network programming language $P4_{16}$. QuIP also comes with the necessary tooling to enable their execution in existing quantum network simulators. We demonstrate its use by showcasing V1Quantum, a completely new device architecture, implementing a link- and network-layer protocol, and simulating it in the existing simulator NetSquid.

## I. INTRODUCTION

QUANTUM communications will enhance our classical (i.e., non-quantum) Internet by supplementing it with capabilities that are provably impossible to achieve purely classically or are significantly more efficient if they can use distributed quantum resources [1]. These new networks will open the door for new possibilities, such as quantum secure communications [2], [3], distributed quantum systems [4],

Manuscript received 30 May 2023; revised 29 November 2023; accepted 20 December 2023. Date of publication 27 March 2024; date of current version 19 June 2024. This work was supported in part by Holland High Tech through Topconsortium voor Kennis en Innovatie [Top Consortium for Knowledge and Innovation (TKI)] Hightech Systemen and Materialen [High-Tech Systems and Materials (HTSM)] (20.0052 Privaat-Publieke Samenwerkingen [Private-Public Partnerships (PPS)]) Funds and in part by Samenwerkende Universitaire Reken Faciliteiten [Co-operative University Computing Facilities (SURF)] Innovation Funds [Wojciech Kozlowski (WK), Rob Smets (RS)]. *(Corresponding author: Wojciech Kozlowski.)*

Wojciech Kozlowski is with QuTech, Delft University of Technology, 2628 CD Delft, The Netherlands (e-mail: w.kozlowski@tudelft.nl).

Fernando A. Kuipers is with the Networked Systems Group, Department of Software Technology, Faculty of Electrical Engineering, Mathematics and Computer Science, Delft University of Technology, 2628 CD Delft, The Netherlands (e-mail: f.a.kuipers@tudelft.nl).

Rob Smets is with SURF, 3511 EP Utrecht, The Netherlands (e-mail: rob.smets@surf.nl).

Belma Turkovic is with TNO, 2595 DA The Hague, The Netherlands (e-mail: belma.turkovic@tno.nl).

Color versions of one or more figures in this article are available at https://doi.org/10.1109/JSAC.2024.3380096.

Digital Object Identifier 10.1109/JSAC.2024.3380096

secure quantum computing in the cloud [5], [6], improved clock synchronisation [7], and quantum-enhanced measurement networks [8]. The idea of a quantum internet has been around for some time [9], [10], but research has significantly accelerated in recent years thanks to large collaborative research efforts [11], [12], [13] and a growing number of testbeds and network deployments in the USA [14], [15], Europe [16], [17], and China [18]. These initiatives explicitly aim to move beyond quantum key distribution (QKD) networks (i.e., quantum networks that can run only QKD) by accelerating research into entanglement distribution networks, thus reaching the next stage in the development of a quantum internet [1].

The field of quantum network protocol research has been growing rapidly over the last few years. Indeed, a lot of networking research is needed as quantum entanglement, the fundamental quantum network primitive, is so different from anything known in classical networks that it does not neatly fit into existing abstractions [10]. In fact, for this reason, at least three quantum network protocol stacks have been put forward [20], [21], [22], [23] and various protocols have been proposed within and outside the context of these new network stacks [19], [22], [24], [25], [26]. While these proposals share many similarities, they often differ in striking ways. For example, some works decouple control data from the quantum payload [19], [22], while others explicitly combine them into quantum packets [26]. Beyond protocols, (quantum) architectures and algorithms for switching, routing, quality of service, and traffic engineering are growing research fields [27], [28], [29], [30], [31], [32], [33], [34], [35].

Rapid progress in hardware capabilities further motivates the need for such work. Various types of quantum end nodes have been demonstrated [36], [37], [38], [39], [40], [41], [42]. Additionally, the number of connected nodes and the achievable distances have also been increasing. The first multi-node quantum network linking quantum processors [38], as well as the first scalable quantum repeater links [43], including one connecting two nodes separated by 33 km [44], have recently been demonstrated in laboratory conditions. However, the timeliness of quantum network protocol research is particularly reinforced by the fact that one of the proposed quantum network stacks has already been demonstrated on real hardware, leading to new insights for their design [45].

The research community has developed several approaches to quantum network architecture and protocol design [22], [24], [25] as well as quantum network simulators to implement and evaluate them, such as QuISP [46], SeQUeNCe [25], and NetSquid [47]. However, the simulators do not share any tooling nor a common node architecture, which has resulted in protocol implementations that are tightly coupled to their simulators. This leads to two key challenges in applying results obtained via such simulations. Firstly, a protocol implemented in one simulator cannot be easily ported and evaluated in another simulator. This results in protocol research happening in silos dictated by the choice of the simulator as it is generally easier to build upon existing results and implementations than porting from another simulator.

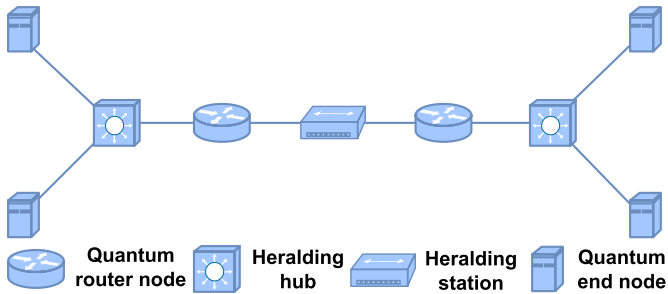

Fig. 1. A sample quantum network. Routers and end nodes generate entanglement between each other via heralding stations and hubs. Routers "stitch" the entanglement to connect distant nodes [9], [19]. The ability to include heralding hubs, nodes with one or more heralding stations with reconfigurable connections, in our networks was a key motivation for QuIP. We considered hubs as a viable strategy to reduce the number of more technologically demanding quantum routers.

Secondly, it makes it difficult to apply prior results and extend existing simulations. Our original research goal was to study control planes for quantum networks with heralding hubs as shown in Fig. 1. A heralding hub is a node with one or more heralding stations and reconfigurable connections and our work was motivated by their potential to reduce the number of technologically demanding quantum routers, making them an attractive prospect for near-term deployments. However, despite the fact that many of the existing protocols do not conceptually preclude such a possibility, all existing implementations were tightly coupled to simulations where all nodes are connected by individual heralding stations. As a result, we spent large amounts of time developing new device and protocol simulation code instead of a control plane, our actual research objective. In addition to the immense time commitment needed for such work, the required low-level quantum device expertise is also non-trivial, making this task only possible for people with a rigorous academic background in quantum technologies.

Therefore, we created QuIP, a P4-based **Qu**antum **I**nternet **P**rotocol prototyping framework. QuIP is a framework for implementing quantum network protocols in $P4_{16}$ [48], a platform-agnostic network programming language. QuIP provides a new abstraction layer that decouples the protocol implementations from simulators. By using QuIP in our own research, we hope to make our results more reusable end extensible. In summary, we make the following contributions:

1) We argue for the use of the domain-specific network programming language $P4_{16}$ for platform-agnostic quantum network protocol implementations.
2) We make our method concrete by providing a set of Python packages enabling its implementation in any Python-based quantum network simulator, such as NetSquid or SeQUeNCe. We provide the concrete integration with NetSquid.
3) We demonstrate the use of QuIP by designing the *V1Quantum* architecture and using it to implement a quantum network protocol stack based on [19] and [22].
4) We open source QuIP [49], [50] and the *V1Quantum* use case study [51], [52].

Note that QuIP is an addition to and not a replacement for any of the existing simulators. Furthermore, since QuIP decouples protocol implementations from simulation, we see potential use cases for QuIP in using of simulations in the development of real-world networks. For example, we have created a proof-of-concept integration of a simulated network running the *V1Quantum* architecture with the orchestration platform of SURF – a national research and education network – and demonstrated it internally.

## II. THE $P4_{16}$ NETWORK PROGRAMMING LANGUAGE

To decouple protocol implementations from simulators, we need a domain-specific language (DSL) and a compiler to translate the DSL into runnable code. In addition to being able to express quantum network protocols, we also had the requirement that the DSL should not assume any specific node architecture. It is our opinion that quantum networks are simply not ready for stable device architectures, and exploring new architectures must remain a crucial part of quantum network protocol research. This statement is motivated by our experience with heralding hubs. Until very recently, such devices have not even been considered in entanglement-based networks [53] and thus they are not accounted for in many of the existing architectures.

Classical programmable networking is an extensive field of research [54], which has resulted in, amongst others, the P4 programming language [55]. P4 is a network programming DSL for expressing the behaviour of the data plane. Despite the fact that it was designed for classical networking, many of its concepts are independent of the nature of the networking unit, be it a packet or an entangled pair of qubits. Indeed, the core concept on which we rely in QuIP is the idea that the data plane is a fixed-function device with programmable blocks within it. The programmer writes P4 programs to specify the behaviour of programmable blocks, but it is the device architecture that defines the interfaces between them and the fixed-function parts of the device. For a more detailed introduction to P4, we refer the reader to [48] and [55].

It has already been shown that a simple entanglement generation protocol running between two neighbouring nodes and a heralding station can be implemented in P4 [56]. Unfortunately, the demo's limited scope masked the difficulty of scaling it up. Firstly, including the quantum physical layer within the P4 program led to a surge in complexity when we started including heralding hubs and routing nodes. Additionally, it coupled the P4 program very tightly to the hardware's low-level design. Furthermore, because entanglement was not a first-class citizen in the device architecture (V1Model), we ended up creating pseudo-architectures where features were being implemented in the user's P4 program, thus coupling the protocol definition to this pseudo-architecture. Ultimately, the core issue was that entanglement is not, and will never be, a first-class citizen in any classical architecture, even when extended with custom external functions as was done in [56]. To promote quantum objects to core abstractions, we needed a pipeline for entanglement that mirrors the one that exists for classical packets. That is, we needed new architectures developed with quantum, and not classical, devices in mind.

Therefore, for QuIP, we chose $P4_{16}$, the second iteration of the P4 language. $P4_{16}$ not only gives flexibility in the programming of the protocols, it explicitly discards the notion of a single generic architecture and allows for vendor-specific architectures and custom interfaces. Just like we did, the P4 community found the single architecture approach insufficient [48]. The authors of [56] did indeed note the architectural flexibility of $P4_{16}$, but they did not fully leverage this feature as they only needed to extend the V1Model with two externs to complete their demo. In this paper, we show how we fully leveraged $P4_{16}$'s ability to support completely custom, but well-defined, device architectures to create QuIP for quantum network protocol prototyping. We showcase its use by designing (Sec. IV) and demonstrating (Sec. V) *V1Quantum*, a new, quantum, $P4_{16}$ target architecture, where entanglement is a core abstraction.

A final, but very important, advantage of P4 is its sizeable ecosystem of open-source tooling such as P4Runtime [57], SDN controllers like ONOS [58], and a compiler [59]. The existence of an active open-source community and its projects significantly reduces the effort that is required to achieve meaningful results. For QuIP, we did not need to implement (and maintain) our own compiler as we were able to use the open-source compiler. Furthermore, it may potentially permit interesting use cases in the future, such as standardising the control- to data-plane interface [60].

## III. Architecture and Implementation

The key challenge in implementing QuIP was that whilst $P4_{16}$ allows for custom architectures, it does not envision adjusting them very often or that researchers would want to propose their own. The idea is that the vendor would provide the hardware implementing a particular architecture, while the programmer would write P4 programs against the architecture's specification. In this section, we explain how we addressed this challenge and implemented QuIP to enable flexible design-space exploration of quantum device architectures. Fig. 2 shows a diagrammatic summary of all the components and their relationships. Later, in Sec. IV, we show QuIP in action through a use-case study. For a guide and demo of QuIP, please see Appendix A.

We separate our discussion into simulator-agnostic and -specific components. In addition to providing a natural implementation boundary, it also represents a boundary between two types of expertise. The simulator-agnostic components are concerned with protocol design against an architecture specification, a task suited to protocol designers who can rely on the abstractions provided by the architecture. In contrast, the simulator-specific components are concerned with the implementation of the architecture in a simulator. This task requires a rigorous understanding of the quantum physics involved in the hardware modelling and careful validation to ensure its accuracy with real devices. This task does not require an extensive understanding of the protocols beyond the device architecture that is meant to be provided.

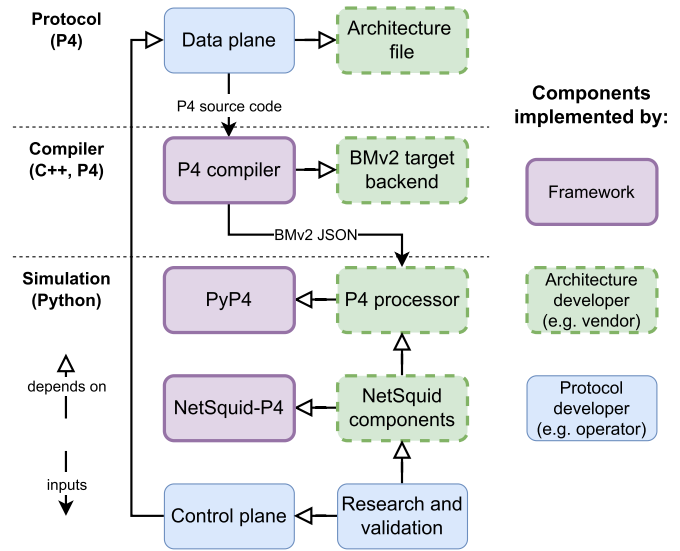

Fig. 2. The software architecture of QuIP. The components are separated into three groups depending on who provides them. Framework: packages that we have developed for QuIP (PyP4, NetSquid-P4) as well as existing tools (P4 compiler). Architecture developer (e.g., vendor): software that needs to be developed for every custom architecture. Protocol developer (e.g., operator): software that the end user of QuIP, i.e., the protocol developer, needs to implement to research and validate protocols.

### A. Simulator-Agnostic Components

*1) Data Plane/Architecture File:* In $P4_{16}$, the programmer writes a data-plane program for a particular architecture. $P4_{16}$ views the data plane as a fixed-function device that has, within it, several programmable blocks [48]. The user writes code to specify the protocols' behaviour within these blocks. The architecture defines the number of blocks, the connections between them, their interfaces, the architecture-specific custom functionality available within, and the surrounding fixed-function processing. By leaving the interface definitions to the architecture, we found $P4_{16}$ sufficiently general to enable quantum architectures. To create a new architecture, the data-plane functionality must be expressed through a P4 architecture file. Note that we are not extending $P4_{16}$ itself, but rather using it as per its existing, classical design and specification.

*2) P4 Compiler/BMv2 Backend:* Once the P4 program, specifying the data-plane protocol and its accompanying architecture file, is ready, it needs to be compiled into some executable format. The open-source reference P4 compiler, p4c [59], supports several backends. For software targets, such as a network simulator, the BMv2 JSON [61] backend is most suitable as the "executable" is expressed as a JSON file. To enable a new custom architecture a new BMv2 backend must be created. Fortunately, since neither the architecture file nor the compiler implement any of the fixed-function parts of the architecture — they only need to know about the programmable blocks, extern function definitions, and their inputs and outputs (including metadata) — adding a new architecture to the BMv2 backend can be done fairly easily.

*3) PyP4/P4 Processor:* Python is the de facto standard programming language in the quantum information community. As a result, most quantum network simulators, such as NetSquid or SeQUeNCe, provide their programming interfaces in Python. Therefore, to execute P4 code in a simulator, we need to have a way of executing it in Python. *PyP4* is a Python package that we provide [49] for implementing P4 processors in a simulator-agnostic fashion. A P4 processor is an object that can be loaded with a BMv2 JSON file produced from compiling a valid P4 source file. It consists of the fixed-function logic that pipes and processes data between the programmable blocks and implements the extern function calls. It will also execute the code within the programmable blocks as per the instructions in the loaded program. *PyP4* also provides the necessary base classes from which the architecture developer can construct their own P4 processor. The architecture developer must also implement the fixed-function logic between the programmable blocks and the extern function calls. For functionality that requires a simulation runtime, a base class must be provided specifying the required interface to the simulation developer. By making the P4 processor simulator-agnostic, we allow for the reuse of P4 programs and architectures with other Python-based simulators, such as SeQUeNCe [25].

*B. Simulator-Specific Components*

For the simulator-specific components, we chose NetSquid [47] as our simulation platform. Primarily for its maturity as compared to SeQUeNCe [25], the other Python-based simulator. We envision that, with QuIP, developers familiar with NetSquid can provide code for a range of architectures, such that protocol developers can use them in their own research, thus saving them the need to implement their own NetSquid simulations from scratch just like we had to.

*1) NetSquid-P4/NetSquid Components:* NetSquid-P4 is a Python package that we provide [50] for architecture developers to develop NetSquid-native interfaces for *PyP4* processors. Examples are available in our *V1Quantum* [51] case study code. Whilst not strictly necessary, we envision that it might be helpful to provide such interfaces in order to facilitate the implementation of the architectures in NetSquid. This is based on our expectation that simulation developers will generally be quantum device experts who are unlikely to be familiar with network protocol design principles or with the P4 language and concepts. Thus, they could otherwise be discouraged from using QuIP due to a learning barrier.

*2) Control Plane:* With P4 we managed to decouple the data-plane protocol implementation from the underlying simulator. However, the control plane remains coupled to NetSquid. In one way, the two planes are already decoupled, because the control plane can populate the data-plane tables by using the interfaces provided by the *PyP4* processor wrapped in *NetSquid-P4*. However, the communication protocol between the two planes remains coupled to NetSquid and its runtime.

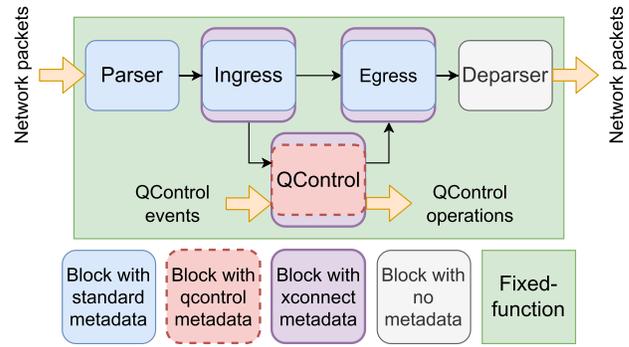

Fig. 3. The *V1Quantum* architecture, based on the V1Model architecture [55]. The classical network path remains unchanged with unmodified standard metadata. *V1Quantum* introduces a new programmable block (QControl), a new type of input (QControl events), and a new type of output (QControl operations). QControl input/output is implemented using a new metadata struct, "qcontrol metadata". Unlike network packets, the QControl events and operations are part of the architecture's fixed-functions. Network packets can be passed to and from the QControl using another new metadata struct, "xconnect metadata".

It is out of the scope of this paper, but it is possible to create a software architecture where the control plane is decoupled from this communication protocol and instead depends only on the table definitions from the P4 program just like it is already done in classical networks with P4Runtime [57].

*3) Research and Validation:* With a new architecture, a researcher can design and implement a new quantum network protocol in P4. However, to evaluate it we still need a device on which the compiled protocol can run. The simulation developer must implement the *device* matching the target architecture. Whilst we had to implement both our own architecture and simulated target device, we are now in a position where the protocol design and simulation can evolve independently of each other. We have found this endeavour fruitful in helping us design useful abstractions for quantum network protocol design, as demonstrated with the *V1Quantum* architecture in Sec. IV. Finally, we found NetSquid's do-it-yourself approach lacking in tools to easily create different topologies, submit requests, record results, and replay simulations. Therefore, we have also implemented NetSquid-Netrunner [62] to aid in creating and running network simulations in NetSquid. Inspired by Mininet [63], it allows the user to quickly create topologies programmatically, configure network demand, run simulations, collect results, and even rerun previous experiments. See Appendix A for more details.

## IV. CASE STUDY: V1QUANTUM

To demonstrate QuIP, we show its use in a case study of our original problem: developing a control plane for a network with heralding hubs (see Fig. 1). To this end, we created *V1Quantum*, an architecture for quantum network devices that we used to construct the entanglement-based network protocols in P4, on top of which we built our control plane. Fig. 3 shows a diagrammatic representation of *V1Quantum*. It was created by extending the V1Model [55], [56] with a new programmable block, *QControl*, that comes with a new type of input, *QControl events*, and output, *QControl operations*. We used QuIP to implement the architecture, limited variations of the link and network layer protocols [19], [22], a limited control plane to reserve entanglement generation paths. Finally, we validated all components in simulation using NetSquid. The code for the *V1Quantum* architecture, including demo code, is available at [51].

### A. Heralding Hubs

In an entanglement-based network, a heralding station is required to connect any pair of routers and/or end nodes. However, despite the fact that there is no fundamental barrier preventing us from combining multiple stations and connecting more than two nodes via reconfigurable optical switches, such "hubs" have not yet been considered in entanglement-based networks. We wanted to study them for two reasons. First, they offered a simpler solution for providing quantum connectivity to a large campus, such as a university. A single "heralding hub" directly connected to multiple end nodes is much easier to deploy than a quantum router connected to each end node via a dedicated heralding station. Furthermore, by virtue of not needing a quantum memory, heralding stations are much simpler to build than quantum routers. Our second reason for considering such networks was that they provide a path of evolution from MDI-QKD (Measurement Device Independent QKD) networks, which are already undergoing trials with industry [64]. MDI-QKD networks are a variant of QKD networks, which, instead of point-to-point connectivity between the end nodes, connect to each other via a central station in a hub-and-spoke network topology. These central stations contain similar hardware to what is required for the heralding stations in an entanglement-based "hub".

We designed *V1Quantum* to support end-to-end entanglement generation protocols based on the QEGP [22] and QNP [19] protocols. However, the key challenge that we faced throughout was not so much in implementing the end-to-end protocol suite itself, but in the inclusion of these heralding hubs in our networks. The QEGP and QNP protocols conceptually had no issue with heralding hubs, but their prior simulated implementations strictly assumed heralding stations. Whilst in the end, we did have to implement all the simulation building blocks ourselves, QuIP allowed us to approach the problem in a two-pronged approach. On one side, we re-implemented simplified variations of the QEGP and QNP protocols in P4. On the other side, we independently implemented the heralding hubs device models in NetSquid. These two pieces are connected via the *V1Quantum* architecture. In the rest of this section, we show how we addressed the various challenges we faced in creating this architecture.

### B. Quantum Network Layers

The problem with the approach of [56] was that it considered the quantum physical layer suitable for implementation in P4. This choice was motivated by the fact that, just like the higher-layer protocols, it included exchanging classical metadata. However, implementing the physical layer in more complex scenarios, such as that of a quantum router, required very tight coupling with the underlying hardware capabilities. Not only did this increase the protocol's complexity, but it also often required updates to the architecture specification for small changes in the physical-layer protocol. The demo's small scale masked the complexity of scaling it up. Therefore, we abandoned this approach and instead opted for an implementation that would more closely resemble the experimental demonstration of [45].

| Layer | Function | Implemention |
|---|---|---|
| *Application* | Quantum network applications | Software (NetSquid) |
| *Transport* | Qubit teleportation | Not implemented |
| *Network (QNP)* | End-to-end entanglement generation | P4 program |
| *Link (QEGP)* | Robust link entanglement generation | |
| *Physical* | Attempt entanglement generation | Fixed-function (NetSquid) |

Fig. 4. *V1Quantum* and the functional allocation protocol stack [22]. This follows the same approach as P4 stacks for classical networks as well as quantum network stack implementations on real hardware [45].

In [45], the authors demonstrated QEGP [22] by implementing the physical-layer protocol in hardware such that it was abstracted away by means of physical-layer commands. Following that approach, we moved the physical-layer protocol to the fixed-function of the architecture and implemented only the link- and network-layer protocols in the P4 data plane as shown in Fig. 4. This approach is also more in line with how P4 is used in classical networks. Reference [22] also defines a transport layer, but we did not need it for our case study.

### C. Quantum Abstractions

With the physical-layer protocol moved to the fixed-function part of the architecture, we proceeded with the implementation of the link (QEGP) and network-layer (QNP) protocols. Here we ran into another problem. The P4 language and the V1Model architecture were designed with classical networks in mind. As a result, all language constructs relate to classical packet processing. We found the lack of abstractions, and thus concrete language constructs, for quantum objects limiting. Reference [56] managed by adding extern function calls. However, externs are limited in the extent to which they can add P4 constructs to represent quantum objects. What we really needed was to elevate entanglement to a first-class abstraction and provide suitable constructs to the programmer directly within the architecture itself. To achieve this, we introduced the QControl programmable block as shown in Fig. 3 with a new type of input and output channels, called QControl events and operations, connected directly to the QControl. We have also coupled the QControl with the classical network Ingress and Egress paths in order to enable the cross-connection of quantum events and operations with classical communication.

To understand the rationale behind this P4 architecture, we first explain the abstractions upon which we built our architecture and data-plane protocols, before discussing its implementation, while also explaining the principles of how quantum networks operate.

*1) Entanglement Objects:* Fig. 5 summarises the principle of operation of our quantum network stack based on [22] and contrasts it with a classical layered network. Fig. 5(a) shows how in a classical network, the payload follows a path from the application at the source node to the application at the destination node. However, in a quantum network, there is no "payload" in the same sense as there is for a classical network. Instead, as shown in Fig. 5(b), the quantum network entangles a qubit at one end node with a qubit located at the other end node, via a chain of intermediate heralding stations and routers. What is then delivered to the application is an *entanglement object*, in our case an entangled pair of qubits, identified through an entanglement identifier that must be identical at both end nodes. Through this shared identifier and its associated entanglement object, the two nodes can identify the qubits in their local quantum memories belonging to the same entangled pair. Operating on entangled qubits is only useful if we can correlate operations on a qubit at one node with operations on the qubit that it is entangled with at the other node. Otherwise, it is impossible to extract any useful information from the entanglement. Additionally, the entanglement object must also contain information about the entangled pair's Bell state. Due to the inherent randomness of quantum mechanics, the network creates one of four random states, called the Bell states, and it is never possible to know ahead of time which one will be created. Without knowing the Bell state, it is also impossible to extract any useful information from the entanglement. Fortunately, while it is impossible to know ahead of time what that state will be for any entangled pair, the network will always be able to infer this information once the entanglement is created [19].

*2) Bell State Measurements:* In addition to the different nature of the payload, Fig. 5 also shows that the way this payload is delivered is quite different. The classical payload begins its journey at the application on the source node, and then by descending and ascending the network layers, it makes its way to the application at the destination node as shown in Fig. 5(a). On the other hand, Fig. 5(b) shows that the qubits that eventually form the entanglement object never leave their nodes. Instead, the qubits at different nodes are entangled with photons via a local process and these photons are sent towards heralding stations where a Bell State Measurement (BSM) is performed on them leading to entanglement between the qubits that were coupled to these photons. At router nodes, a local BSM is performed (usually called an entanglement swap in this context) that, similarly to the BSM at the heralding station, leads to entanglement between the remote qubits with which these two local qubits were entangled. Taking Fig. 5 as an example, performing a BSM at node 3 on qubit 3.1 and 3.2 leads to entanglement between qubits 1 and 5. Therefore, we see that BSMs "stitch" long-distance entanglement together from shorter entanglement objects. The heralding station BSM combined two pairs, each consisting of a local qubit entangled with a photon, into a single entangled pair between the qubits at the two nodes surrounding the station. The router BSM combined two link-level entangled pairs into a single end-to-end entangled pair. Thus, we also see that entanglement

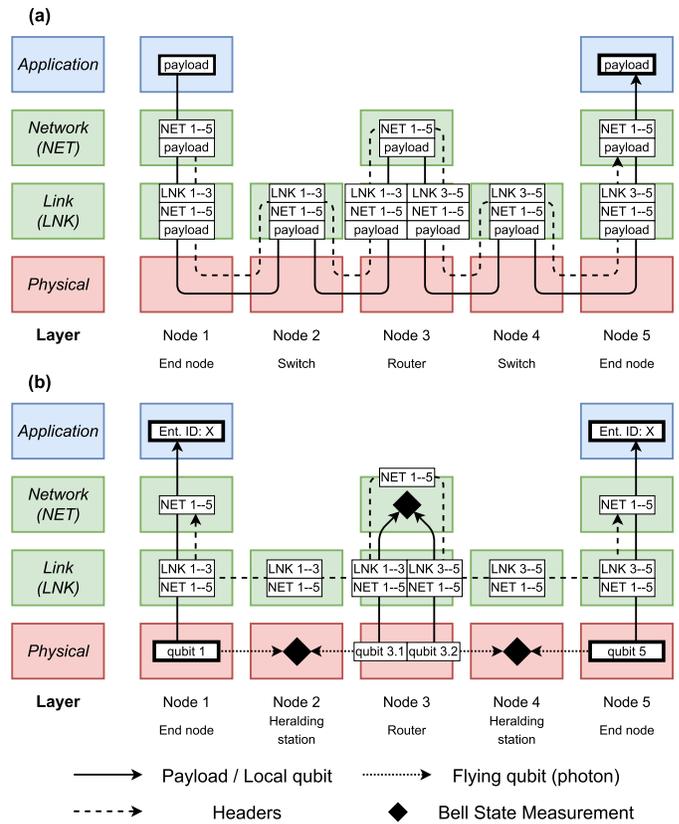

Fig. 5. (a) The path across the network of a classical payload and its packet headers. In a classical network, the packet descends and ascends through the network layers, losing and gaining headers as it makes its way from source to destination. The link header identifies the source and destination on a local network and the network header identifies the network source and destination. (b) The multitude of paths that qubits and headers take in our protocol implementations to deliver a single end-to-end entangled pair of qubits. Unlike in the classical case, the "payload", i.e., the entangled qubits "bubble up" from the physical layer to "surface" at the application layer rather than beginning their journey from there. Here, the link header maps a link-local entanglement object to its qubits and the network header maps the end-to-end entanglement object to its qubits at the end nodes.

"bubbles up" from the bottom of the network stack as it is extended via successive BSMs. As a BSM is a measurement of the state of two qubits, it will produce an output containing two bits of the measurement outcome, one for each qubit. This outcome is used to identify which of the four Bell states was produced through the BSM. Additionally, as not all physical BSM implementations are deterministic (in particular optical implementations, such as the one used at heralding stations, are not deterministic), the BSM event must also indicate the success or failure of the attempt.

*3) Classical Communication:* A key feature of this entire process is that at each layer, classical messages must be exchanged in order to correctly deliver an entanglement object to the next layer. At either the physical or link layer, the heralding station must send a signal back to the nodes that originated the photons. In [22], this was done at the physical layer, but in our implementation, we did it via the link layer (although *V1Quantum* also sends the signal on the physical layer). Going higher, the network-layer protocol must send messages to identify the chain of all the link-layer entanglement objects that were combined via BSMs at intermediate routers so that the two end nodes can unambiguously identify which qubits at the two nodes belong to the same entanglement object. These messages must also collect all the outputs of the BSMs, which contain information about the entangled pair's Bell state [19].

### D. Implementation

*1) QControl:* The first element of our V1Quantum implementation is the QControl programmable block as shown in Fig. 3. The block is part of the target architecture and thus it is defined, including its inputs and outputs, in the P4 architecture file. There are three key features of this new block: (1) a new input channel that triggers within the QControl block without a parser, (2) a new output channel that leaves from the QControl block without a deparser, and (3) it is connected to the classical Ingress and Egress paths.

*a) QControl events:* It follows from the previous discussion that the QControl block must receive the BSM measurement outcomes. Therefore, we introduced a new input channel, QControl events. Via this channel, the QControl receives the outcomes from both the heralding station as well as the entanglement swap BSMs. The events do not need to be parsed as they are part of the fixed-function definition and can be immediately delivered in a structured form as "qcontrol metadata" to the program. The QControl events are defined by the target architecture, but the processing of the events, i.e., the data-plane protocol logic, is implemented in the P4 program. However, the events themselves originate from the target device as shown in Fig. 3. See Appendix B for details.

*b) QControl operations:* The QControl block must also issue commands to perform these BSMs. To issue the entanglement swap BSM we introduced a new output channel, QControl operations. Via this channel, the QControl can issue commands which will be executed by the fixed-function. Implementing these operations via a new output channel is more aligned with P4's packet processing, where operations on the payload, such as copying, dropping, or forwarding, are performed by the fixed-function rather than via extern calls. However, we have implemented the heralding station BSMs differently, because they require the photons from the nodes to arrive at precisely the same time. The alternative mechanism is described in Sec. IV-D.3. QControl operations do not need deparsing, because similarly to the input channel they are implemented via the "qcontrol metadata" in the program. Also similarly to the events, the QControl operations are defined by the target architecture and they are emitted as a result of the data-plane protocol logic implemented in the P4 program. Once emitted the operations are executed by the target device as shown in Fig. 3.

*c) Cross-connect path:* Finally, in order to build the chain of link-layer entanglement objects and collect all the BSM outcomes required to infer the final Bell state, the QControl block needs to be able to receive and send classical packets. Therefore, the QControl programmable block is connected to both the Ingress and Egress paths. This allows the QControl to consume metadata about entanglement from a remote node and to send such metadata itself. The cross-connect path is also defined in the target architecture with an implementation on the target device. However, since the classical packet pipeline does not depend on any quantum behaviour, which would require simulation, it is implemented in the simulator-agnostic *PyP4* processor and thus can be re-used across different simulators.

*2) Entanglement Headers:* To implement the QEGP and QNP protocols, we needed to express the entanglement objects and classical messages sent by those protocols using P4 constructs. P4 headers were the natural choice for this task. Combined with P4 registers to store state information locally on each node they also proved to be sufficient and we were able to implement our concept of network-level entanglement objects as explained in Sec. IV-C directly within the P4 program itself.

Whilst these protocols did not consider the classical messages as headers, by implementing them in P4, we found that a limited analogy can exist as illustrated in Fig. 5(b). Firstly, indeed, the headers do not travel with the payload like they would in a classical network. They have to remain decoupled as the qubits that end up forming the final entanglement object never physically leave their nodes. However, just like classical headers identify the source and destination of the packet at their respective layer, we defined our quantum headers such that they identify the end-points and the state of the entanglement object at their respective layer. That is, we implemented the headers such that the link-layer header identified the qubits and their state at the nodes on either side of the link while the network-layer headers identified the qubits and their state at the end nodes. Thus, at every layer, the header identifies an entanglement object, but the higher the layer, the larger the scope of the entanglement.

However, whilst the idea of identifying the end-points is similar, the principle of operation deviates from the classical one. For example, it is not possible for one node to fill out all the entanglement object information in the header. Indeed, the network layer collects this entanglement object information by using the link-layer headers to follow the chain of entanglement from one end node to the other. As shown in Fig. 5(b), the network layer has a link-layer header pushed on top before being sent to the next router. At the router, the link-layer header uniquely identifies which qubit was used in the local BSM that "stitched" this link entanglement object with the next link. The network-layer header picks up the corresponding link-layer header and continues along the path. Once it reaches the remote end node, the final link-layer header is used to identify the end node's qubit that belongs to the end-to-end entanglement object identified by the network-layer header. Additionally, at each layer we have two headers for each entanglement object moving in counter-propagating directions. The counter-propagating message is needed to ensure that both nodes receive acknowledgement that the full chain of BSMs has been achieved and that none of the intermediate pairs were dropped due to losses, such as decoherence [19].

*3) BSM Groups:* The QControl block allowed us to move the physical-layer protocol into the fixed-function section,

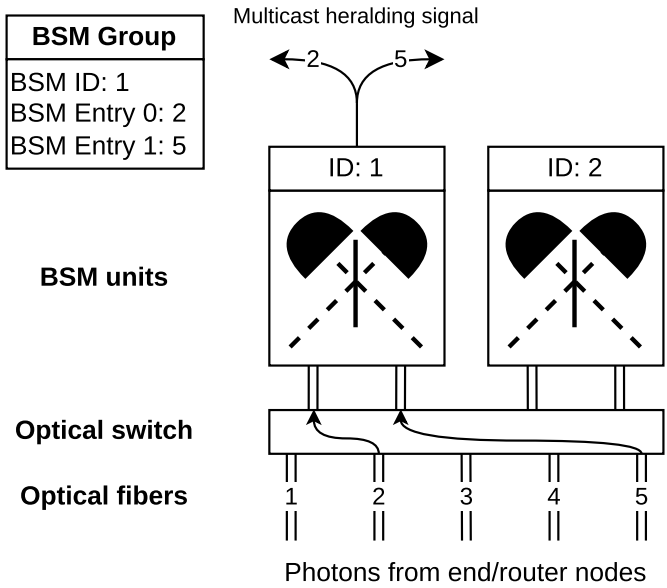

Fig. 6. A BSM group at a heralding hub. Installing a BSM group will connect a pair of fibres, identified by the BSM entries, to a BSM unit, identified by the BSM ID, and initiate entanglement generation. The BSM's output is available in the P4 program and multicast at the physical layer to the nodes connected via that particular unit. Entanglement generation continues until the group is removed.

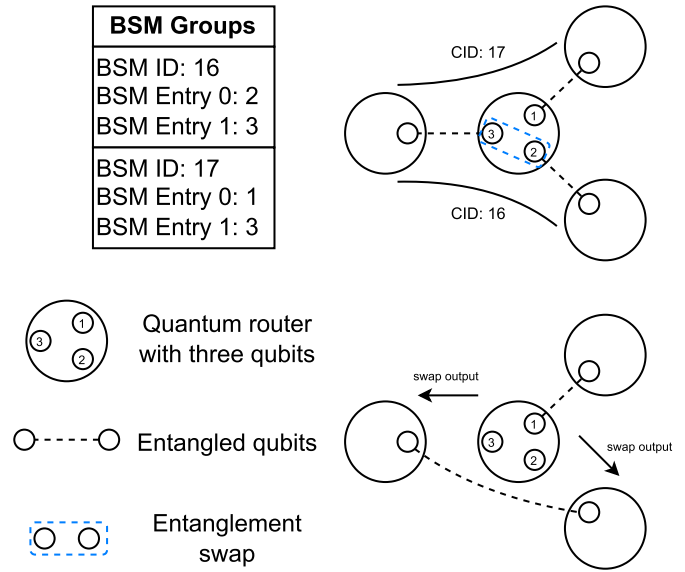

Fig. 7. A BSM group at a quantum router. Entanglement swaps are triggered and assigned to connections through a connection ID (CID) using match+action rules. In parallel, we install a BSM group with the BSM ID equal to the CID and its two BSM entries matching the connection's upstream and downstream ports. Once the BSM completes, the result is multicast using the swap's CID as the BSM ID.

thus simplifying the implementation of an end-to-end protocol suite. However, alone this does not enable the use of heralding hubs. We considered hubs composed of one or more sets of BSM units, each capable of performing a BSM on its two optical inputs, and an optical switch that would connect the fibres from the nodes to the BSM units. While a hub with multiple BSM units has not yet been realised for entanglement generation, they are technologically feasible.

To achieve the desired control over these hubs, we needed:

1) The ability to program the connections from the incoming fibres to the right detector unit.
2) The ability to trigger entanglement generation between the nodes connected via this unit.
3) The P4 data-plane protocols to have access to the detector output.

For the first point, it was sufficient if this was only accessible to the control plane, which is analogous to how classical networks reconfigure their connections. Therefore, we introduced BSM groups, which are exposed by the device, but not in the P4 program, for the control plane to configure. A BSM group consists of three identifiers: the BSM ID, which identifies the detector unit, and two entries, which identify the input fibres. Upon creation of a BSM group at a heralding hub, the device was expected to switch the input fibres identified by the group's entries to the BSM unit identified by the BSM ID. The BSM output, the heralding signal, is then automatically multicast, at the physical layer, i.e., not via the P4 data plane, to the nodes at the remote ends of the fibres. Physical-layer multicasting is required to implement protocols in alignment with the experimental demonstration of [45]. This is illustrated in Fig. 6.

For the second requirement, we chose to deviate from the original QEGP design [22] by making the BSM group's creation trigger entanglement generation between the two nodes connected by these input fibres. In the original design, the heralding station's BSM unit was always connected to the same two nodes. Entanglement generation was triggered by these two nodes according to a queue that they maintained in a distributed fashion. Whilst possible to extend this approach to multiple nodes connected via a central hub, we opted to instead have the hub trigger entanglement generation to avoid problems due to queue convergence.

To implement the last requirement, the BSM output is also copied to the P4 data plane at the heralding hub as a QControl event. Furthermore, the multicast signal, upon arrival at the end node or router, is also immediately forwarded to the node's P4 data plane as a QControl event. This allows us to implement a link-layer protocol with its logic fully contained at the router and end nodes, just like [22], but it also allows for implementations with some logic implemented at the hub. Indeed, we used this ability in our QEGP variation to have the hub assign unique labels to each entangled pair on the link. These labels are later used at the other nodes to identify which end-to-end entanglement object they belong to. To achieve this, we needed the QControl to be able to multicast a packet to the two nodes connected to the particular BSM unit. Therefore, when the control plane creates a BSM group, in addition to connecting the fibres to the BSM unit, it will also create a special multicast group, this time available in the P4 program, with the same ID as the BSM group and connecting the same two nodes.

We also used the BSM group abstraction for the entanglement swap BSM performed at the routers (Fig. 7). However, here we did not need the BSM group to establish any physical connectivity between the qubits like it had to for fibres at the heralding hub. Nevertheless, we still found the BSM group's multicasting functionality useful. After an entanglement swap, the QNP protocol has to multicast the BSM result to the next upstream and downstream nodes. We implemented the QNP in *V1Quantum* by leveraging BSM groups for this purpose. More concretely, for each network-layer connection, we create a corresponding BSM group at each router along the path with the BSM ID matching the connection ID and the BSM entries identifying the upstream and downstream ports. Additionally, the router's data plane is programmed to use the connection ID as the BSM ID for the entanglement swap BSM operation. Thus, when it completes, the QControl is able to use this ID to multicast the outcome using the appropriate BSM group.

### E. Limitations of V1Quantum

To facilitate the proof-of-concept implementation of the *V1Quantum* architecture, we made two simplifying assumptions: (1) Routers have only one qubit per link and each qubit is uniquely assigned to that link only. That is, the qubits cannot be used to generate entanglement with another node via a link that it is not assigned to. (2) Qubits are assumed to never be kept in memory long enough for decoherence to have visible effects. In other words, *V1Quantum* assumes that the qubit memory lifetime is long compared to the time it takes to generate end-to-end entanglement. These are not fundamental limitations and should be resolvable with further extensions to *V1Quantum* using QuIP. Additionally, all classical connections in our simulations were considered reliable and in-order. This is a reasonable assumption given that TCP was used for the implementation of the QEGP protocol in the lab [45]. This suggests that the classical network ports in quantum architectures should be logical rather than physical. Instead, it is up to the architecture to define and the user to configure the properties of these connections.

## V. EVALUATION

We evaluated QuIP on a network with a heralding hub running the protocol suite we implemented on the *V1Quantum* architecture. The evaluation code can be found online [65].

### A. Demand

As no large-scale quantum network exists yet, we do not have any real demand patterns to base our simulations on. For our evaluation, we opted for a demand pattern that we considered realistic for near-term QKD. QKD works by first producing a finite-size block of raw key material. In entanglement-based networks, the raw key material is obtained by generating a fixed number of entangled pairs and measuring the qubits at the two end nodes. The measurement results form the raw key material. For our evaluation, we focused on the raw key generation step as it is the only step that involves entanglement, and we do not simulate any of the classical post-processing that is usually required. Thus, in our simulations entanglement is generated on demand in response to requests issued by individual end nodes for a fixed number (50) of entangled pairs with some other end node submitted at random intervals at a pre-specified network-wide average rate. For each request, the two end nodes are chosen at random. One of them is chosen to be the requesting node and is responsible for forwarding the request to the remote end node.

### B. Topology

We use a heralding-hub-and-spoke topology. We vary the number of BSM units at the hub between one and eight. As each BSM unit can only handle one request at a time, their number determines the maximum throughput. All our simulations are performed on a hub-and-spoke topology with 16 end nodes. Each end node is connected to the central hub by 5 km of fibre, a reasonable distance for a small network.

### C. Control Plane

The control plane we used for this proof-of-concept evaluation is simple: its only goal is to ensure that all requests eventually complete and it is not optimised for any metric. It runs on a controller located at the hub. All requests are submitted by both end nodes to the controller, which schedules them and, at the scheduled time, configures the data plane. Requests are scheduled on a first-come-first-served basis and run until completion before their resources can be re-assigned. Since multiplexing entanglement generation on a single link is not currently possible, any fibre can only be assigned to one request at a time. Therefore, the controller will skip requests if the required fibres are occupied. Whilst BSM units can also only be assigned to one request at a time, they can be flexibly reconnected to different links via BSM groups. Therefore, although they impose a limit on the number of requests that can be handled simultaneously, they do not block requests based on which fibres are required. More advanced control planes are a subject of ongoing research and may be developed in the future.

### D. Physical Parameters

All the physical parameters and code used for the simulation are available online [65] with instructions on how to reproduce the results of our evaluation. In summary, we use realistic values for fibre losses and photon detector properties. We model the stationary qubits at the end nodes based on Nitrogen Vacancy quantum nodes [42].

### E. Results

We simulated randomly generated demand for different network-wide request rates ten times for 2 seconds each. Fig. 8 shows averages calculated based on requests that started and completed between 1 s and 2 s (i.e., once an equilibrium is reached). Fig. 8(a) shows that our control plane easily saturates the hub's resources in terms of throughput (completed requests per second). However, Fig. 8(b) shows that this happens at the cost of mean latency (time between the moment a request is submitted and the moment the request starts being executed). This happens even if the network is far from being saturated. Furthermore, the variation in latencies is large, especially when the network is close to saturation, as seen in Fig. 8(c), which shows the cumulative distribution function of completed requests for the case of 6 BSM units. For reference, a request of 50 entangled pairs needs only ≈0.015 s to actually execute.

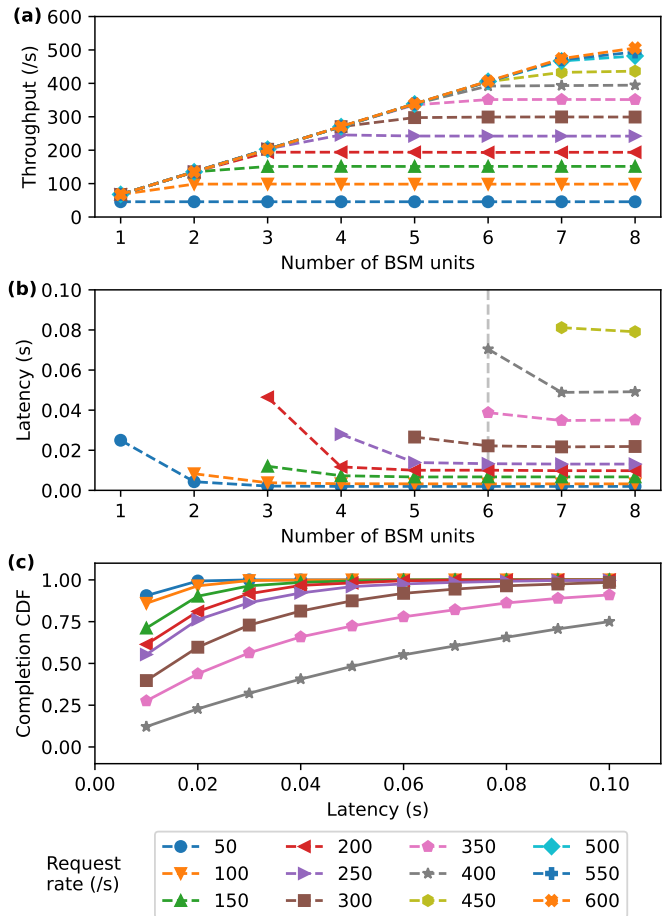

Fig. 8. The performance of a heralding-hub-and-spoke network. Different coloured lines correspond to a different network-wide request rate. (a) The mean throughput vs. the number of BSM units. We see that our protocols easily saturate throughput. Error bars are not shown as they are of the same order of magnitude as the markers. (b) The mean latency vs. the number of BSM units. As the request rate increases so does the mean latency. The number of BSM units itself does not have much impact on latency when significant bandwidth is available. The variation in latency is enormous and thus, instead of error bars, (c) shows the cumulative distribution function of completed requests for 6 BSM units. We see that as the request rate increases, the distribution significantly spreads out.

## VI. CONCLUSION

We have introduced QuIP, a P4-based Quantum Internet Protocol prototyping framework. QuIP decouples quantum network protocol implementation from simulation by using the $P4_{16}$ network DSL. We demonstrated the framework in a case study in which we introduced *V1Quantum*, a quantum device architecture where entanglement is a core abstraction. We showed how we used $P4_{16}$'s architectural flexibility to incorporate heralding hubs and implement a quantum network protocol suite. We provide the Python packages that we developed for QuIP to enable its integration with Python-based quantum network simulators and provide an integration with NetSquid. We have also made our code publicly available [49], [50], [51], [62].

## APPENDIX A
## QUICK START GUIDE

Whilst QuIP requires many components, it is not expected that all of them are implemented by a single party. A natural division of responsibilities would have protocol experts develop the data- and control-plane protocols, while quantum experts implement the simulated devices. The architecture will ideally be developed jointly as it defines the interface between the two domains.

### A. Simulator-Agnostic Components

*1) Data Plane/Architecture File:* The first step is to define the device architecture in a P4 file. A device architecture defines the processing pipeline of the device by defining the user-programmable blocks, their inputs and outputs, the architecture-specific functionality available within the blocks as well as the fixed-function processing that is executed between the blocks. The reference P4 compiler [59] includes several architectures in the `p4include` directory that can be used as a starting point. The V1Quantum architecture file can be found in the `p4include` directory in our fork of the P4 compiler [52]. Once an architecture is expressed in a P4 file, data-plane protocols can be written for it. The P4 source files for the protocols we implemented for the V1Quantum architecture can be found in the `v1quantum/protocol/data_plane/p4` directory of the V1Quantum repository [51].

*2) P4 Compiler/BMv2 Backend:* Once ready, the P4 source files must be compiled into a format that can be executed in a simulator. QuIP uses the BMv2 JSON format [61] for this purpose. Thus, for every architecture a P4-to-BMv2 compiler is required. The easiest way to create one is to fork the open-source P4 compiler [59] and add the new architecture to it. The compiler frontend is shared by all architectures and since $P4_{16}$ supports custom architectures by default, there is no need to provide new frontend code. However, a new BMv2 backend must be added to `backends/bmv2`. Fortunately, backend code for BMv2 is relatively straightforward since it compiles to a JSON file rather than a hardware-specific binary format. Furthermore, since the compiler does not implement any of the fixed-function logic, it only needs to know how to convert programmable blocks, externs, and metadata blocks into JSON structures. This work is quite mechanistic and a developer with moderate C++ familiarity should be able to adapt code from an existing BMv2 backend. The backend code for the V1Quantum architecture can be found in `backends/bmv2/quantum_switch` in our fork of the P4 compiler [52].

*3) PyP4/P4 Processor:* To execute a BMv2 JSON file, a processor is required. For QuIP simulations, the processor is expected to be implemented in Python. For the purposes of implementing such a processor, QuIP provides the *PyP4* package [49]. *PyP4* provides functions for processing the BMv2 code so that the developer does not need to understand the BMv2 format themselves. However, the JSON file does not contain any information on how the blocks are connected nor any architecture-specific fixed-function logic, including externs. Therefore, the architecture developer must additionally provide the following three elements to implement a processor: (1) the connections between the inputs and outputs between the different programmable blocks, (2) the fixed-function logic executed by the data plane between the programmable blocks, and (3) the entry points and the logic for all the externs that are called from within the programmable blocks. For fixed-function logic that cannot be implemented in a simulation-agnostic manner the architecture developer can instead provide abstract base classes that define the interfaces that must be implemented within the simulator. The V1Quantum processor implementation can be found in the V1Quantum repository [51] in the file `v1quantum/processor.py`.

### B. Simulator-Specific Components

*1) NetSquid-P4/NetSquid Components:* Finally, the architecture needs to be implemented in the quantum network simulator. This implementation must provide all the fixed-function capabilities specified by the architecture. In the case of quantum network devices this will include functionality such as the physical-layer protocol for entanglement generation and entanglement swapping. For V1Quantum we implemented two devices: a quantum node, and a heralding hub. The code can be found in `v1quantum/components` of the V1Quantum repository [51]. To facilitate this task, the architecture developer may choose to provide a simulator-native interface for the P4 processor from the previous step. The purpose of such an interface is to facilitate the integration of the P4 processor by providing functions that can accept simulation-native objects, cast them to architecture-specific metadata constructs, inject them into the P4 processor, cast any output metadata into simulation-native objects, and call back into the simulation for further execution. This is optional, but it may help if the simulation developer is different from the protocol and architecture developer. For V1Quantum we have implemented the `v1quantum/device.py` of the V1Quantum repository [51] using the QuIP package *NetSquid-P4*.

*2) Control Plane:* The task of a control plane is to populate the match+action tables defined in the P4 data plane such that the desired networking objectives are achieved. As the tables, matches, and actions are defined in the P4 source code, they are all available in the compiled BMv2 JSON file. Therefore, a P4 processor implemented using *PyP4* will automatically provide the necessary API calls for inserting and removing entries to and from these tables such that they can be used by the processor. QuIP makes no assumptions as to whether the control plane is distributed or centralised. The control plane that we implemented for the case study in this paper is available in `v1quantum/protocol/control_plane` in the V1Quantum repository [51].

*3) Research and Validation:* With a device architecture, a P4 data plane, and a control plane we are ready to simulate them in a network scenario and evaluate their performance. We recommend using the *NetSquid-Netrunner* [62] package that we have created for QuIP to construct and run such simulations. The package provides modules to easily generate and instantiate network topologies as well as generate and inject user demand. The V1Quantum repository [51] includes two examples of experiments created using *NetSquid-Netrunner* in the `experiments` directory. In particular, the `experiments/hub` experiment was used to produce the data in the case study in this paper. For a step-by-step example that executes the simulations we used to produce the data in this paper, please follow the `README.md` in [65].

## APPENDIX B
## V1QUANTUM ARCHITECTURE IN P4

The new QControl input and output channels, called QControl events and operations, are implemented with a new metadata block called "qcontrol_metadata", which is defined as:

```
@metadata @name("qcontrol_metadata")
struct qcontrol_metadata_t {
  QControlEventType event_type;
  bit<64> event_timestamp;

  // QControl operations and parameters.
  // Note that currently port == qubit.
  QControlOperation operation;

  // QControlOperation == release.
  bit<9> release_qubit;

  // QControlOperation == swap.
  bit<16> swap_bsm_id;
  bit<9> swap_qubit_0;
  bit<9> swap_qubit_1;

  // QControl event metadata.

  // Heralding/swap BSM outcome.
  bit<16> bsm_id;
  bool bsm_success;
  bit<2> bsm_bell_index;
}

enum QControlEventType {
  heralding_bsm_outcome,
  swap_bsm_outcome,
  cnetwork
}

enum QControlOperation {
  none,
  swap,
  release
}
```

When the QControl block is triggered, the triggering event is specified in the `event_type` field of `qcontrol_metadata_t`. The other fields of the metadata struct will be filled in depending on the type of the event. There are three types of events:

- `heralding_bsm_outcome`: A heralding signal notifying of success or failure and the resulting state of link-level entanglement between two nodes connected via the heralding station. This event is triggered at the heralding hub, but also at the router and end nodes as the heralding signal is also transmitted at the physical-layer protocol.
- `swap_bsm_outcome`: An entanglement swap has completed at a quantum router node as a result of the entanglement `swap` operation (which is also a BSM). The QControl metadata will indicate the ID that was assigned to the operation and the resulting Bell state.
- `cnetwork`: The QControl was triggered by a classical message that was diverted from the Ingress block. In this case, the remaining fields `qcontrol_metadata_t` are not filled in and the program should inspect the diverted packet fields.

The QControl block issues operations to the local quantum network processing unit by filling out `qcontrol_metadata_t`. It indicates the operation type by setting the `operation` field and sets parameters by filling out the accompanying fields. There are three types of operations:

- `none`: Do nothing. This is useful if QControl cannot act on the provided input, it only needs to record some metadata locally in registers, or it only wants to emit a classical packet.
- `swap`: Execute an entanglement swap (BSM) operation. The QControl must assign an ID to the operation, which will be returned with the entanglement swap result as a QControl event, and choose the two qubits that are to be entanglement swapped.
- `release`: Release the qubit so that it can be reused. This is equivalent to a packet drop in classical networking. This operation is extremely useful in handling window conditions that result when the network's state changes and has not yet converged into the new equilibrium.

The QControl also has the ability to send and receive classical packets. It is coupled to the classical networking path by means of the "xconnect_metadata" defined as:

```
@metadata @name("xconnect_metadata")
struct xconnect_metadata_t {
  // Choose pathway for the packet.
  PathWay pathway;

  // Ingress information.
  bit<9> ingress_port;

  // Send single packet.
  bit<9> egress_spec;

  // BSM group multicast.
  bit<16> bsm_grp;

  // BSM info available at egress.
  bit<16> bsm_info;
}

enum PathWay {
  cnetwork,
  qcontrol
}
```

If Ingress wants to divert a packet to QControl, it must set the `pathway` field to `qcontrol`. If the QControl wants to send out a classical packet, it can either set the `egress_spec` field, which has exactly the same meaning as the `standard_metadata_t` field of the same name or, if a BSM group has been installed, it can set the `bsm_grp` field to multicast a packet as per the definition of the BSM group (see Sec. IV-D.3). The `ingress_port` indicates the ingress port of the packet and `bsm_info` is filled out by the BSM group for use in the Egress block.

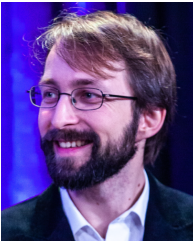

**Wojciech Kozlowski** received the M.Sci. degree in theoretical quantum physics from the University of Cambridge in 2012 and the D.Phil. degree in atomic and laser physics from the University of Oxford in 2016. After graduating from Oxford, he decided to change track and he worked as a Software Engineer with the Network Software Team, Metaswitch, for two years. This has placed him in the unusual position of having experience in both computer networking and quantum physics, which is how he ended up with QuTech, The Netherlands. There, he worked as a Post-Doctoral Researcher with the Stephanie Wehner Group for two years, before changing roles. In his new role as a Quantum Network Engineer, he works on developing the software and network protocol stacks for the Quantum Internet Alliance in which he leads the Stack and Integration Team.

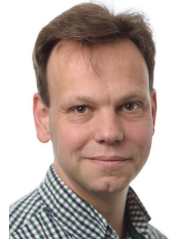

**Rob Smets** joined the Dutch Organization for Applied Scientific Research (TNO), where he acted as an Innovator in many mobile and fixed-network projects in 2008. After completing the Ph.D. degree, he worked on all-optical signal processing with Philips Research, he started working with Lucent Technologies in 1999, where he as a Member of Technical Staff, supported the Optical Networking Group and the Access Network Division, Alcatel-Lucent. In 2013, he joined the Dutch National Research and Network Operator, SURF, as a Network Architect, and an Innovator, with a focus on introducing new network technology and services. At present, he has taken an interest in the cutting-edge between scientific and operational aspects in the field of DevOps, quantum information networking, and network services.

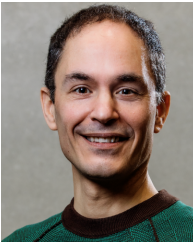

**Fernando A. Kuipers** (Senior Member, IEEE) received the Ph.D. degree (cum laude) from Delft University of Technology (TU Delft) in 2004. He was a Visiting Scholar with the Technion—Israel Institute of Technology in 2009 and Columbia University, New York, NY, USA, in 2016. He is currently a Full Professor with TU Delft, where he established and leads the Networked Systems Group and the Laboratory on Internet Science. He co-founded the Do IoT Fieldlab and the PowerWeb Institute. His research interests include network optimization, network resilience, quality-of-service, and quality-of-experience and address problems in computer networks, software-defined networking, 6G, and the Internet of Things. His work on these subjects includes distinguished papers at IEEE INFOCOM 2003, Chinacom 2006, IFIP Networking 2008, IEEE FMN 2008, IEEE ISM 2008, ITC 2009, IEEE JISIC 2014, NetGames 2015, and EuroGP 2017. He has served as the General Chair and the TPC Chair for flagship conferences, such as ACM SIGCOMM (2021 and 2022) and IEEE INFOCOM (2024). He is the Vice Chair of the ACM SIGCOMM Executive Committee. He served on the Board of the TU Delft Safety and Security Institute. He is the Co-PI of the Dutch 6G Flagship Project Future Network Services, where he leads the program line Intelligent Networks.

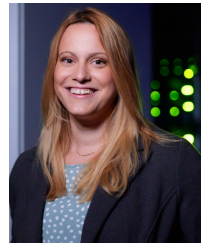

**Belma Turkovic** received the M.Sc. degree in electrical engineering from the University of Sarajevo, Bosnia and Herzegovina, in 2015, and the Ph.D. degree from the Embedded and Networked Systems Group, Delft University of Technology, The Netherlands, in 2022. Since 2021, she has been a Research Scientist with the Dutch Organization for Applied Scientific Research (TNO), where her research focuses on the development of 5G/6G core networks and applications utilizing them, cloud computing, programmable infrastructure (IaC), and programmable networks (e.g., programmable data planes and software-defined networking).